\documentclass[prb,twocolumn,aps]{revtex4-1}
\usepackage{graphicx}
\usepackage{dcolumn}
\usepackage{bm}
\usepackage{hyperref}
\renewcommand{\vec}[1]{{\mathbf{#1}}}
\newcommand{\beq}{\begin{eqnarray}}
\newcommand{\eeq}{\end{eqnarray}}

\begin{document}\title{ Non-conservation of Fermionic Degrees of Freedom at Low-energy in Doped Mott Insulators}
\author{\small Shiladitya Chakraborty }
\author{Seungmin Hong}
\author{Philip Phillips}
\affiliation{Department of Physics,
University of Illinois
1110 W. Green Street, Urbana, IL 61801, U.S.A.}
\date{\today}

\begin{abstract}
    Hall and optical conductivity experiments on the cuprates indicate that the low-energy fermionic degrees of freedom in a
    doped Mott insulator posess a
    component that is dynamcially generated and hence determined by
    the temperature.  We show explicitly how the spectrum in the lower Hubbard band should be partitioned to describe such dynamically generated charge degrees of freedom and corroborate this picture with the results from the exact low-energy theory of the Hubbard model.
   A consequence of such dynamics is that the Landau one-to-one correspondence between bare
    electrons and the effective fermionic degrees of freedom at low energies breaks down explicitly. This state of affairs obtains because the total hole number
    is not conserved as it contains a dynamical contribution. We propose that any
    experimental probe that couples to the low-energy
    dynamics of a doped Mott insulator, quantum oscillation
    experiments included,
   should be interpreted in terms of the total dynamically generated hole
   number rather than the bare value. 
\end {abstract}
 
\maketitle
 
In 1993, Meinders, Eskes, and Sawatzky\cite{meinders} concluded based on an exact diagonalization study that because the
effective number of low-energy degrees of freedom in a doped Mott
insulator is a function of the hybridization and therefore the volume
and temperature, ``...it is not possible to define a Hamiltonian that
describes the low-energy-scale physics unless one accepts an effective
nonparticle conservation.''  Particle non-conservation as used here
refers to the fact that the number of low-energy degrees of freedom is not
strictly determined by the electron filling or equivalently the doping level
but rather by dynamical degrees of freedom generated from the
hybridization and hence the temperature. If this statement is correct, then it
must be the case that the chemical potential for the static fermionic low-energy degrees of freedom in any realistic model for a doped Mott insulator
is not equivalent to that of the conserved charge, namely the bare
electrons. Thus far, an explicit construction demonstrating this has
not been advanced.  Given the obvious importance of this result, it is
surprising how little attention it has attracted.  In this paper, we
directly address the question of how particle conservation breaks down
in a low-energy theory of a doped Mott insulator.  We first show that
experiments on the Hall\cite{nishikawa,ando,ono,gorkov} and optical\cite{cooper2,cooper,uchida1} conductivities and general
theoretical considerations support this claim.  Finally, we propose a
simple partitioning of the spectral weight in the lower Hubbard band
(LHB) which isolates the explicit hybridization-dependent degrees of
freedom that are responsible for the dynamical generation of charge
degrees of freedom and hence effective particle
non-conservation as defined above.  We show that these degrees of freedom can be
understood within the recently\cite{ftm1,ftm2,ftm3,ftm4,ftm5} derived exact low-energy theory of the
Hubbard model. 

\section{Experimental Motivation}

Is there any experimental indication in doped Mott systems that the
number of charge carriers is dynamically generated?  It would suffice
to show that either 1) the carrier density is temperature dependent or
2) the number of charge carriers exceeds the nominal doping level,
hereafter referred to as $x$.  Consider first the experiments on the
Hall coefficient in La$_{2-x}$Sr$_x$CuO$_4$ (LSCO).  In the underdoped
regime, the inverse Hall number is strongly temperature
dependent\cite{nishikawa,ando,ono}.  Gor'kov and
Teitel'baum\cite{gorkov} observed that a two-component empirical formula, 
\beq\label{n} 
n_{\rm Hall}(x,T)= n_{0}(x) + n_{1}(x)\exp(-\Delta(x)/T),
\eeq
accurately describes the inverse Hall coefficient in LSCO in the
underdoped regime.  One of the
components is independent of temperature, n$_0(x)$, given by the
static doping level, while the other is
strongly temperature dependent, $n_{1}(x)\exp(-\Delta(x)/T)$.  
The key observation here is that the temperature dependence in
$n_{\rm Hall}$ is carried entirely within $\Delta(x,T)$ which defines a
characteristic activation energy scale for the system.  Gor'kov and
Teitel`baum's\cite{gorkov} analysis suggests that the activation energy is set by the pseudogap energy
scale.  Consequently, the bound component should be liberated beyond
the $T^*$ scale for the onset of the pseudogap. 

Additionally, optical conductivity experiments indicate that the
effective number of charge carriers exceeds the nominal count provided
by the doping.  In the experiments\cite{cooper,uchida1} the integrated optical
conductivity,
\beq\label{neff}
N_{\rm eff}(\Omega)=\frac{2m V_{\rm cell}}{\pi e^2}\int_0^\Omega
\sigma(\omega)d\omega
\eeq
is generally plotted in which the cutoff is set by the optical gap, which for LSCO is
$\Omega\approx 1.2eV$.  Here $\sigma(\omega)$ is the optical conductivity, $V_{\rm cell}$ the
unit-cell volume per  formula unit, $m$
the free electron mass, and $e$ the electron charge.  In a rigid-band
semiconductor model in which spectral weight transfer is absent,
$N_{\rm eff}=x$.  However, in all cuprates, regardless of whether they are electron\cite{cooper} or hole-doped\cite{cooper2,uchida1}, $N_{\rm eff}$
exceeds $x$ in the underdoped regime. 

Consequently, experimental
probes which couple to the current reveal that the number of charge
carriers in the cuprates is 1) temperature dependent and 2) exceeds
the nominal doping level, consistent with Meinders, et al\cite{meinders}.
An interesting question is how does one define the chemical potential for such dynamically generated charge
degrees of freedom.  Clearly it is not equal to that of the
bare electrons as the the effective number of charge degrees of
freedom exceeds the bare charge count.   We argue below that the
effective doping level that captures the dynamical generation of the
charge degrees of freedom as in Eq. (\ref{n}) is given by
\beq
x'=x+\alpha
\eeq
where $\alpha$ is a dynamical correction determined by the
hybridization. This redefinition of the doping level naturally
arises from the exact\cite{ftm1,ftm2,ftm3,ftm4} low-energy theory of the Hubbard model which has
been shown to explain\cite{ftm1,ftm3,ftm5} both Eq. (\ref{n}) and  Eq. (\ref{neff}).

\section{Redefinition of Chemical Potential}
 
The goal in this section is to redefine the chemical potential so that
the effective number of fermionic charge carriers is consistent with dynamical
generation of charge degrees of freedom discussed in the previous section.  In the standard theory of metals, the intensity or spectral weight of a band is
completely exhausted by counting the number of electrons it can hold. That is, 
it is a constant given by one per unit cell and per spin direction.  Essential 
to this view is the robustness of electron quasiparticles even in the presence 
of interactions.  Because of the one-to-one correspondence between
electrons and quasiparticles, the chemical potential, $\mu$, can be defined
either by counting electrons
\beq\label{chemelec}
n=\int_{-\infty}^\mu N(\omega)d\omega,
\eeq
or by integrating,
\beq\label{chemhole}
y=\int_\mu^{\Lambda_g} N(\omega)d\omega,
\eeq
the unoccupied part of the spectrum.  Here $N(\omega)$ is 
the single-particle electron density of states and $\Lambda_g$ is a
cutoff demarcating the low-enegrgy physics.  As a result of the
electron-quasiparticle correspondence, $y$ is identical to the number
of doped holes, $x$, and the electron filling is given
by $n=2-x$ (for a single band).  

In stark contrast, the empty part of the spectrum per spin at low energies, Eq.
(\ref{chemhole}), exceeds the doping level in strongly correlated systems such
as doped Mott insulators. The inherent problem with strongly
correlated systems is that the energy bands are not the traditional
static bands that typify band insulator systems.  This can be illustrated simply by considering
the Hubbard model
\beq\label{hubb}
H_{\rm Hubb}
	&=&	-t\sum_{\langle i,j\rangle,\sigma}  c^\dagger_{i\sigma}c_{j\sigma}
		+U\sum_{i} c^\dagger_{i\uparrow}c^\dagger_{i\downarrow}c_{i\downarrow}c_{i\uparrow},
\eeq
in which electrons hop among a set of lattice sites, but pay an energy cost $U$ 
whenever they doubly occupy the same site. Here  $i,j$ label lattice 
sites, $\langle i,j\rangle$ indicates nearest neighbors, $c_{i\sigma}$ 
annihilates an electron with spin $\sigma$ on site $i$ and $t$ is the 
nearest-neighbor hopping matrix element. When $t=0$, the Hamiltonian
is diagonal
\beq
H_U	= U\sum_i n_{i\uparrow}n_{i\downarrow}
	= \frac{U}{2}\sum_{i\sigma}\eta^\dagger_{i\sigma}\eta_{i\sigma},
\eeq
where $\eta_{i\sigma}= c_{i\sigma} n_{i\bar\sigma}$ creates the 
excitations above the gap in the upper Hubbard band (UHB) on sites occupied by a single electron.  Its complement, $\xi_{i\sigma}=c_{i\sigma}
(1-n_{i\bar\sigma})$ creates excitations strictly on empty sites and hence 
describes particle motion below the gap.  Here $\bar\sigma=-\sigma$. Consequently, the anticommutator 
 \beq
{\rm m}^0_{\rm LHB}
	= \frac{1}{N}\sum_{i,\sigma} \langle\{\xi_{i\sigma},\xi_{i\sigma}^\dagger\}\rangle=2-n,
\eeq
determines the spectral weight in the lower Hubbard band (LHB). Since 
each hole in a half-filled band decreases the single occupancy by one, the weight of the UHB is $1-x$.  Because the total weight of the
UHB and LHB must be 2, we find that $2-n+1-x=2$ or $n=1-x$ and ${\rm
  m}^0_{\rm LHB}=1+x$ in the atomic limit. The weights $1+x$ and $1-x$ also determine 
the total ways electrons can occupy each of the bands. Thus, in the atomic limit, electrons 
alone exhaust the total degrees of freedom of each band. Further, since each hole 
leaves behind an empty site that can be occupied by either a spin up or a spin 
down electron, the electron addition spectrum in the LHB has weight 
$y=2x$\cite{meinders,eskes,stechel}.  Hence, the occupied part of the LHB and UHB both 
have identical weights of $1-x$ in the atomic limit. 

Because the operators $\xi$ and $\eta$ do not diagonalize the hopping term, the 
total intensity of the LHB 
\beq\label{hl}
{\rm m}_{\rm LHB}
	=1+x+\frac{2t}{U}\sum_{ij\sigma}g_{ij}
		\langle \tilde c_{i\sigma}^\dagger \tilde c_{j\sigma}\rangle+\cdots
	=1+x+\alpha,\nonumber\\
\eeq
has $t/U$ corrections as shown by Harris and Lange\cite{hl}. Here 
  $\tilde c_{i\sigma}$ are related to the original bare fermion operators via a 
canonical transformation that brings the Hubbard model into block diagonal 
form in which the energy of each block is $nU$. In fact, all orders of 
perturbation theory\cite{eskes} increase the intensity of the LHB beyond its 
atomic limit of $1+x$. It is these dynamical corrections that $\alpha$ 
denotes.  While the intensity of the LHB increases away from the atomic limit, 
the total number of ways of assigning electrons to the LHB still remains fixed 
at $1+x$. Consequently, the total weight of the LHB exceeds the fermionic phase space 
and additional degrees of freedom are needed. 

\begin{figure}
\includegraphics[width = 8.0cm]{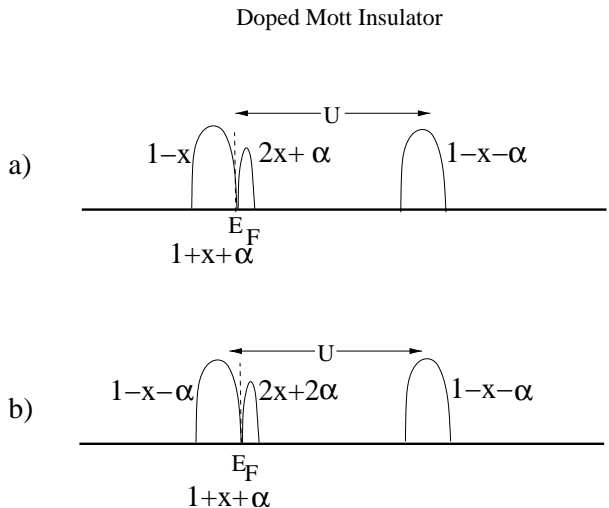}
\caption{Redistribution of spectral weight in the Hubbard model upon doping the
		insulating state with $x$ holes. $\alpha$ is the dynamical correction 
		mediated by the doubly occupied sector. To order $t/U$, this correction 
		worked out by Harris and Lange\cite{hl}. a) The traditional approach
		\cite{meinders,randeria,stechel} in which the occupied part of the 
		lower band is fixed to the electron filling $1-x$.  b)
                New assignment of the spectral weight in terms of
                dynamically generated charge carriers.  In this
                picture, the weight of the empty part of the LHB per spin is
                the effective doping level, $x'=x+\alpha$. }
\label{spec}
\end{figure}

Nonetheless, the sum of the spectral weights in the LHB and UHB must be 2 by charge conservation. 
Consequently, the weight in the UHB, ${\rm m}_{\rm UHB}=1-x-\alpha$, decreases 
faster than $1-x$. How should the spectrum in the LHB be partitioned? Harris and Lange\cite{hl} did not address 
this issue possibly because when the total weight of a band exceeds the electron 
count, the chemical potential for the effective low-energy degrees of
freedom is not simply understood within a conventional picture of
adding or removing electron quasiparticles.  Despite this difficulty, it is common 
in the strongly correlated
community\cite{meinders,eskes,stechel,randeria} to assign the spectral
weight for the bare electrons assuming the doping level is not
renormalized by the dynamics.  Hence, the weight below the chemical
potential, $n$, remains\cite{meinders,eskes,stechel,randeria} at 
$1-x$ and the weight immediately above the chemical potential becomes $y=2x+\alpha$ 
as depicted in Fig.\ (\ref{spec}a). On this account\cite{meinders,eskes,stechel,randeria}, only the states above the 
chemical potential acquire doubly occupied character dynamically. 

This assignment of the chemical potential is valid for the bare
electrons alone and does not include the dynamically generated
charges as distinct low-energy entities. To address this problem, consider the
Lehmann representation,
\beq\label{lehman}
m_{\rm LHB}
	&=&\displaystyle\sum_{k,m}\displaystyle\int_{-\infty}^{\mu}\kern-.5emd\omega
		|\langle\psi_m^{N-1}|c_k|\psi_g^N\rangle|^2\delta(\omega-E_m^{N-1}+E_g^N)\nonumber\\
	&\kern-3.5em+&\kern-2em
		\displaystyle\sum_{k,m}\displaystyle\int_\mu^{\Lambda_g}\kern-.5emd\omega
		|\langle\psi_m^{N+1}|c^\dagger_k|\psi_g^N\rangle|^2\delta(\omega-E_m^{N+1}+E_g^N)
\eeq
of the spectral function. In these expressions, 
$E_m^N$ is the $m^{\rm th}$ eigen-energy of the N particle system with ground state 
$E_g^N$ with associated many-body states $|\psi_m^N\rangle$ and $|\psi_g^N\rangle$, 
respectively.  The filled (first term) and empty (second term) parts of the spectrum do not have traditional definitions in terms of $n$ and $1-n$ respectively if one insists on setting $n=1-x$.  That is, $x$ is not a
fundamental property of the spectral function as it is for a
rigid-band system such as a band insulator in which it equals the empty part of the spectrum. The disconnect between $x$ and the empty part of the
spectrum in the LHB (the second term in Eq.\ (\ref{lehman})) obtains because
holes are generated either by doping or mixing with the
doubly occupied sector.  This is not an option for a band insulator. As a result, the number of holes is
 {\it not} determined strictly by the doping.  That is, although $n$ is well-defined, $1-n$
 does not have any fundamental meaning in terms of the integrated
 spectral weight of the empty part of the LHB. In fact, the empty part of the spectrum has no
 obvious relation to anything.

We now show how the spectrum can be partitioned so that the chemical
potential accounts for a charge number consistent with Eq. (\ref{n}). Note we have some degree of freedom in describing the physics in the LHB since it is not a rigid band.  If the dynamical contribution can be removed through a re-definition of the chemical potential, then the empty part of the
spectrum per spin will be the effective hole number. The justification for this picture is as follows.
 In a hole-doped system, turning on 
a finite $t/U$ creates pairs of double occupancies and empty sites (doublon-holon pairs).  The weight in the UHB corresponds to adding one electron in the  high 
energy sector, in other words creating double occupancy. Doublon-holon 
pairs clearly deplete this intensity leading to a loss of spectral in
the UHB faster than the atomic limit value of $1-x$. The occupied weight in 
the LHB corresponds to removing an electron in the low-energy sector. In other 
words, the occupied part of the spectrum corresponds to removing an electron such 
that the number of double occupancies remains conserved. Hence the occupied part 
of the LHB is a measure of single-occupancy whose weight as well must decrease 
on creation of  doublon-holon pairs. In other words, the weights in the occupied part 
of the LHB and the UHB must be the same, since both provide a  measure of the 
same phase space. Therefore, we propose that the consistent definition of the 
chemical potential for the low-energy fermionic degrees of freedom can
be obtained by demanding that the two weights be equal.  Note this says nothing about the nature of the excitations which live in the high-energy scale. Consequently, we arrive at the 
assignments of the spectral weights in Fig.\ (\ref{spec}b).  The occupied part of the LHB has weight 
($1-x-\alpha$) and the unoccupied part 
$2(x+\alpha)$.  The fermionic degrees of freedom that are
associated with this assignment of the chemical potential reflect the
dynamical generation of the charge degrees of freedom.    As a
result of the dynamics,  $x'=x+\alpha$ now denotes the effective 
number of hole degrees of freedom per spin at low
energy. Consequently, we propose that it is with respect to $x'$ that a Luttinger 
theorem exists not $x$, the bare hole number.

In the case of 
electron doping, the chemical potential ($\mu$) lies in the UHB  where  $2x$ 
electron removal states are created below $\mu$ and the weight above $\mu$ 
is given by $1-x$ in the atomic limit. Turning on a finite $t/U$ creates 
doublon-holon pairs. In this case, the holes belong to the LHB and represent 
the high-energy configurations of the system.  The weight above $\mu$ represents 
the amplitude  for adding an electron to the UHB, or creating a double 
occupancy, which is depleted upon creation of doublon-holon pairs since neither 
holons nor doublons can contribute to the creation of double occupancies upon 
addition of a single electron.  This weight is analogous 
to that of the occupied part of the LHB in the case of hole
doping.  For charge-transfer systems, such as the cuprates, the same argument applies because
of the equivalence\cite{meinders} with the Hubbard model for realistic values of the
hybridization between the bands.

To counter the argument that the dynamical corrections might not
affect the physics on all energy scales, it suffices to compute the
 cross correlator between $\xi_{i\sigma}=c_{i\sigma}(1-n_{i\bar\sigma})$ and $\eta_{i\sigma}=c_{i\sigma}n_{i\bar\sigma}$. The full 
electron spectral function, $A(\vec k,\omega)=-{\rm Im}FT(\theta(t-t')\langle
\{c_{i\sigma}(t), c^\dagger_{j\sigma}(t')\}\rangle)/\pi = A_{\eta\eta}+A_{\xi\xi}
+2A_{\eta\xi}$, contains two diagonal terms  $A_{\eta\eta}$ and $A_{\xi\xi}$ and 
a cross term $A_{\eta\xi}$ which represents the degree to which the high and low 
energy degrees of freedom are coupled.  Here, FT represents the frequency and 
momentum Fourier transform.   We have computed $A_{\eta\xi}$ previously\cite{slavery} 
and it is clearly non-zero at all frequencies that bracket the turn-on of the 
spectral weights in the LHB and UHB at half-filling and at finite doping. This is 
simply a reflection of the fact that at all frequencies, the states in the LHB all 
have doubly occupied character.  The dynamical contribution reduces the 
spectral weight.  Let us call the reduction $q$ and hence the weight 
is given by $1-x-q$.  The weight in the unoccupied part of the LHB is $2x+\alpha+q$. 
For the weight of a hole per spin to be equal to that of an electron, we must have 
that $q=\alpha$. This results in the assignments in Fig.\ (\ref{spec}b).  

As a result, the bare electrons and the low-energy dynamically generated
fermionic charge carriers in the LHB do not stand in a one-to-one correspondence.  The efficient cause of this breakdown is
 dynamical spectral weight transfer. Insertion of an electron affects
the spectrum at all energies while only local changes occur in terms
of the low-energy degrees of freedom.  Such an orthogonality
catastrophe is due entirely to the
existence of the UHB\cite{slavery,anderson} and persists as long as the
degrees of freedom transferred from the UHB provide a relevant
perturbation to those in the LHB. In fact, Fig. 1b provides a possible basis for the Anderson\cite{anderson} conjecture
that the very existence of the UHB (in the form of dynamical spectral weight transfer) leads to a breakdown of Fermi
liquid theory.
\begin{figure}
\includegraphics[width = 8.0cm]{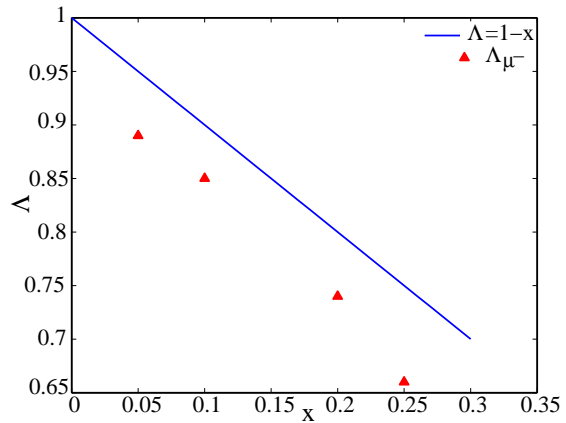}
\caption{Integrated  spectral weight in the occupied part of the lower 	
		Hubbard band, $\Lambda_{\mu^-}$, from the charge 2e low-energy 
		theory\cite{ftm1,ftm2,ftm3,ftm4} with $U/t=8$.  Here $x$ is the doping level for the 
		conserved charge, $Q=\sum_{i\sigma}c_{i\sigma}^\dagger c_{i\sigma}+
		2\sum_i\varphi^\dagger_i\varphi_i$.  Clearly shown is that the occupied 
		part (red triangles) of the one-particle spectrum has a weight less than 
		$1-x$ (solid blue line).  }
\label{intsum}
\end{figure}

An experimental prediction of this work is that $\alpha$ should be temperature dependent.  Making contact with Eq. (\ref{n}), $\alpha$ should turn on at $T^\ast$.  As a result, the dynamical part of the spectral weight signifies an opening of the pseudogap in the single-particle spectrum as pointed out earlier\cite{ftm1,ftm5}. This is reasonable for  two reasons. First, if $\alpha\ne 0$, the number of ways of adding a particle exceeds the number of ways of adding an electron to the empty part of the spectrum in the LHB.  That is, some of the particle addition states in the LHB are orthogonal to the addition of an electron. Second, there is no reason to separate the UHB and LHB's if there is no gap between them.  Consequently, the collapse of the UHB should be coincident with the closing of the pseudogap. Recently, Peets, et al.\cite{peets} have observed that the UHB collapses once the pseudogap closes, consistent with our prediction here.

\section{Confirmation from Exact Low-Energy Theory}

Since the weight of the band in which the chemical potential resides in a doped Mott 
insulator exceeds the electron count, new degrees of freedom are required in any 
consistent low-energy theory. The extra degrees of freedom are generated from mixing 
with the doubly occupied sector and hence should emerge upon integration of the states 
far away from the chemical potential.  We have carried\cite{ftm1,ftm2,ftm3,ftm4} out 
this Wilsonian program exactly for the Hubbard model and showed that a charge $|2e|$ 
bosonic field emerges. The boson which is non-propagating has charge 2e for hole 
doping and -2e for electron doping, represents the mixing with double occupancy and 
double holes respectively.  For hole doping, the conserved charge $Q$, which equals the total electron filling $n$\cite{ftm1,ftm2,ftm3,ftm4},
 is a sum of  two components,
\beq\label{Q}
Q=\sum_{i\sigma}a_{i\sigma}^\dagger a_{i\sigma}+2\sum_i\varphi^\dagger_i\varphi_i,
\eeq
immediately implying that the weight of the fermionic part must be
less than the conserved charge. Here $a_{i\sigma}$ is the annihilation operator for the fermionic degree of freedom that results when the high-energy scale is integrated out and $\varphi$ is a charge $2e$ boson. That 
$Q$ is the conserved charge
can be verified by inspection as it trivially commutes with the
low-energy effective Hamiltonian. In fact, Eq. (\ref{Q}) gives a prescription for $\alpha$,
namely the bosonic charge, if we interpret $Q$ as $1-x$ and the
fermionic quasiparticle density as $1-x'$.  In this theory\cite{ftm1,ftm2,ftm3}, the quasiparticles 
are transformed at low energies to 
\beq\label{cop}
c^\dagger_{i,\sigma}
	\rightarrow (1-n_{i,\bar\sigma})c_{i,\sigma}^\dagger 
		+ V_\sigma \frac{t}{U} b^\dagger_i c_{i,\bar\sigma}
		+ V_\sigma \frac{t}{U}\varphi_i^\dagger c_{i,\bar\sigma},
\eeq
to leading order in $t/U$ upon the integration of the high energy scale. Here 
$b_i = \sum_j b_{ij} = \sum_{j\sigma}c_{j\sigma}V_\sigma c_{i\bar\sigma}$ 
with $V_\uparrow = -V_\downarrow = 1$ and $j$ a nearest-neighbour of site $i$. The first two terms represent the standard 
electron operator in the lower Hubbard band dressed with spin fluctuations which constitutes the quasiparticles or the effective fermionic degrees of freedom. However, 
the last term represents the correction due to dynamical spectral weight transfer. 
Eq.\ (\ref{cop}) lays plain that an electron at low energy contains a propagating 
part that arises from the charge 2e boson. To illustrate that more than just the 
fermions are needed to satisfy the $1-x$ sum rule, we computed the pure fermionic part of the spectral function by evaluating 
the Green function: $\int d\varphi^*d\varphi FT(\int[Dc_i^*Dc_i]c_i(t)c_j^*(0) 
\exp(-\int L_{\rm IR} dt))/Z$. $L_{\rm IR}$ is the low-energy 
Lagrangian\cite{ftm1,ftm2,ftm3,ftm4} obtained by integrating out the UHB and $Z$ the 
partition function.  We computed this quantity assuming that the boson is spatially 
homogeneous, which is justified\cite{ftm1,ftm2,ftm3,ftm4} since there are no gradient 
terms of the boson in the low-energy action. The results in Fig.\ (\ref{intsum}) 
demonstrate that the integrated weight in the occupied part of the
spectrum is indeed less than $1-x$.  That this weight is less than
$1-x$ is independent of the approximations used to calculate the
spectral function.  This follows entirely from the fact that the conserved
charge, $Q$ is a sum of a fermionic and a bosonic part. The deficit from $1-x$ is carried 
in the $\varphi^\dagger c_{i\bar\sigma}$ term.  The difference between
the red triangles and the solid line approximates $\alpha$.

\section{Concluding Remarks}

Experimentally, any measurement which probes the fermionic low-energy degrees of 
freedom should be interpreted in terms of the total number of hole degrees of freedom, 
$x+\alpha$ not $x$. For example, the superfluid density should 
exceed $x$ and scale as
as $x+\alpha$, already confirmed in
YBa$_2$Cu$_3$O$_{6+x}$\cite{cooper2} (YBCO).  Similarly, Fermi surface volumes, that is the total volume of the hole pockets minus that of the
electron pockets,
extracted from quantum 
oscillation experiments\cite{osc}, whose origin is still not understood, should be 
compared with $x'$ not $x$ as the experimental probe is the current.  This is particularly germane because
the Fermi surface
volumes extracted experimentally\cite{osc,osc1} for YBCO are not consistent with any integer
multiple of the physically doped holes. Interestingly, the first
experiments of this type observed oscillations in the Hall
coefficient\cite{osc}.  Hence, it is perfectly reasonable that the
effective doping level should be consistent with the physics that leads to Eq. (\ref{n}).   

Finally, Fermi liquid theory is recovered when the charge 2e boson 
 decouples from the electronic spectrum. By decoupling we mean that the UHB collapses and the LHB has a weight of $2$. In this limit, there is no true high-energy scale and $\varphi$ should be an irrelevant degree of freedom. To illustrate, using the appropriate\cite{scal}
scaling such that the kinetic energy remains constant in the
limit $d\rightarrow\infty$, that is, $t\rightarrow O(1/\sqrt{d})$, and
averages of the form $\langle c^\dagger_j c_i\rangle\propto
1/\sqrt{d}$ (note as $d\rightarrow \infty$ the scaling of $c$ and $\tilde c$ (Eq. (\ref{hl})) are not
trivially related), we find
that the boson-dependent terms in the exact low energy theory, $t\sum_i\varphi_i^\dagger
c_{i\uparrow}c_{i\downarrow}\rightarrow O(1/\sqrt{d})$, $t^2/U\sum_i\varphi_i^\dagger\varphi_i\rightarrow O(1/d)$ and
$t^2/U\sum_{\langle ij\rangle} \varphi_i^\dagger b_{ij}\rightarrow
 O(d\times(1/\sqrt{d})^3)$, vanish when $d=\infty$.   Consequently, no breakdown of Fermi liquid theory obtains
 as seen numerically\cite{scal} for $d=\infty$ and $n\ne 1$.  
In finite dimensions, the precise value of the coupling
constant and doping level at which the bosonic degrees of freedom
decouples remains the open problem in Mottness.

\acknowledgements This work was funded by the NSF DMR-0605769, NSF
PHY05-51164 (KITP) and
Samsung scholarship (S.\ Hong).  We also thank F.
Kr\"uger, T. Stanescu, R. Leigh, and G. Sawatzky for critical remarks.

\end{document}